 \documentclass[final,5p,times,twocolumn]{elsarticle}

 \usepackage{graphicx}

\usepackage{amssymb}
\usepackage{amsthm}

\usepackage{footnote}
\usepackage{dcolumn}
\usepackage{sidecap}
\usepackage{threeparttable}
\usepackage{hyperref}

\biboptions{sort&compress,numbers}

\begin{document}

\begin{frontmatter}



\title{First observation of the competing  fission modes in the neutron-deficient sub-lead region }


\author[cenbg]{I. Tsekhanovich\corref{cor1}}
\cortext[cor1]{Corresponding author}
\ead{tsekhano@cenbg.in2p3.fr}

\author[york,jaea,isolde]{A.N. Andreyev}

\author[jaea]{K. Nishio}

\author[cenbg]{D. Denis-Petit}

\author[jaea]{K. Hirose}

\author[jaea]{H. Makii}

\author[frib]{Z. Matheson}

\author[riken]{K. Morimoto}

\author[riken,kyushu]{K. Morita}

\author[frib]{W. Nazarewicz}

\author[jaea]{R. Orlandi}

\author[vecc,hbni]{J. Sadhukhan}

\author[riken,kyushu]{T. Tanaka}

\author[jaea]{M. Vermeulen}

\author[lublin]{M. Warda}

\address[cenbg]{CENBG, CNRS/IN2P3-Universit\'e de Bordeaux, 33170 Gradignan, France}

\address[york]{Department of Physics, University of York, York YO10 5DD, United Kingdom}
\address[jaea]{Advanced Science Research Center, Japan Atomic Energy Agency, Tokai, Ibaraki 319-1195 Japan}
\address[isolde]{ISOLDE, CERN, CH-1211 Geneve 23, Switzerland}

\address[frib]{Department of Physics and Astronomy and FRIB Laboratory, Michigan State University, East Lansing, Michigan 48824, USA}

\address[riken]{RIKEN Nishina Center for Accelerator-Based Science, Saitama 351-0198, Japan}

\address[kyushu]{Department of Physics, Kyushu University, Fukuoka 819-0395, Japan}

\address[vecc]{Variable Energy Cyclotron Centre, Kolkata 700064, India}
\address[hbni]{Homi Bhabha National Institute, Mumbai 400094, India}

\address[lublin]{Institute of Physics, Maria Curie-Sk{\l}odowska University, 20-031 Lublin, Poland}

\begin{abstract}

Fragment mass distributions from fission of excited compound nucleus $^{178}$Pt have  been deduced from the measured fragment velocities. The $^{178}$Pt nucleus was created at the JAEA tandem facility in a complete fusion reaction $^{36}$Ar + $^{142}$Nd, at beam energies of 155, 170 and 180 MeV. The data are indicative of a mixture of the mass-asymmetric and mass-symmetric fission modes associated with higher and lower total kinetic energies of the fragments, respectively. 
The measured fragment yields are dominated by asymmetric mass splits, with the symmetric mode contributing at the level of $\approx1/3$. This constitutes the first observation of  a multimodal fission in the sub-lead region.
Most probable experimental fragment-mass split of the asymmetric mode, $A_{L}/A_{H}\approx 79/99$, is  well reproduced by nuclear density functional theory using the UNEDF1-HFB and D1S potentials. 
The symmetric mode is associated by theory with very elongated  fission fragments, which is consistent with the observed total kinetic energy/fragment mass correlation.

\end{abstract}


\end{frontmatter}


\section{Introduction}
\label{introduction}

Understanding of the nuclear fission process is important for many areas of fundamental science, technology, and medicine. In particular, fission   is crucial for the existence of many transuranium nuclei, including the predicted long-lived superheavy isotopes \cite{Oganessian2015}, as well as for the heavy element formation in the astrophysical $r$-process \cite{Goriely2013,Thielemann2017,Giuliani2018,Eichler2015}. 
Better knowledge of fission properties is also essential for our understanding of the antineutrino flux from nuclear reactors \cite{An2017,Giunti2017}. 
Regardless of the area, one needs detailed information on fission rates and fission fragment (FF) mass distributions (FFMDs). 

At present, our experimental knowledge of fission is primarily limited to nuclei close to the stability line \cite{Andreyev2018,Hessberger2017} and within a fairly narrow isospin range $N/Z\sim1.48-1.58$. Extrapolation of this knowledge to higher  neutron-excess regions ($N/Z>1.8$) relevant to the $r$-process is highly model dependent \cite{Goriely2013,Giuliani2018,Eichler2015}. 
While there has been exciting progress in global modeling of nuclear properties, facilitated by advanced computing, a comprehensive, microscopic explanation of nuclear fission is still difficult to achieve, due to complexity of the process \cite{schunck2016,krappe2012}.
To advance theoretical modeling of fission, experimental FFMDs data are needed in broader range of $N/Z$-values, to test the isospin dependence of model predictions.

Due to its experimental accessibility, the neutron-deficient  sub-lead region ($N/Z\sim1.3$) provides excellent testing ground for studies  of the isospin dependence of fission observables.  
Due to exotic $N/Z$ ratio, new facets of fission process can be expected.
Indeed, the observation of asymmetric fission of $^{178,180}$Hg \cite{Andreyev2010,Liberati2013} attributed to shell effects in pre-scission configurations 
 \cite{Moller2012,Warda2012,McDonnell2014,Ichikawa2018} has generated an appreciable interest  in this region, both experimentally and theoretically.
Inspired by the $^{180}$Hg results,  FFMDs have been experimentally studied  for several neutron-deficient sub-lead nuclei \cite{Liberati2013,Nishio2015,Prasad2015,Tripathi2015}. 
As shown by theory \cite{Andreev2012,Moller2012,Ichikawa2012,Warda2012,Andreev2013,McDonnell2014,Ghys2014}, the topology of potential energy surfaces (PES) in sub-lead nuclei is significantly different (flat, broad and rather structureless) from those in the actinides,  which explains fairly low dependence of the corresponding experimental FFMDs on the compound nucleus (CN) excitation energy (cf. \cite{Nishio2015}). According to the global survey of FFMDs  \cite{Moller2015},
a new extended region of asymmetric fission is expected  in neutron-deficient Re-Pb isotopes with $98\leqslant N\leqslant 116$. It is 
separated from predominanly asymmetrically-fissioning actinides
 by a zone of symmetric fission around Ir-At in the vicinity of $N\sim$ 120--126 \cite{Andreyev2018}, whose properties were extensively investigated in the past (cf., e.g., Refs.~ \cite{Itkis1985,Itkis1991}).
The experimentally studied neutron-deficient $^{178,180,182,190,195}$Hg and $^{179,189}$Au isotopes \cite{Andreyev2010,Liberati2013,Nishio2015,Prasad2015,Tripathi2015} lie on the northern border of this region. As concluded in Ref.~\cite{Moller2015}, new high-quality FFMDs data for selected sub-lead isotopes are needed to test and guide theoretical developments.

In the transitional regions between asymmetrically and symmetrically fissoning sub-lead nuclei, an interplay between different fission modes
might exist, by analogy to light  \cite{Schmidt2000,Bockstiegel2008} and  heavy \cite{Hulet1986, Hulet1989} actinides. In view of PES properties in the sub-lead region \cite{Andreyev2010,Moller2012,Warda2012,McDonnell2014}, 
an observation of a competition between fission modes will shed light on the nature of   near-scission configurations of nuclei, which are some 60 nucleons lighter and greatly deficient in neutrons, as compared to actinides and transactinides.  
This Letter provides the first experimental demonstration of the existence of competing fission modes in sub-lead  nuclei, by revealing the presence of   asymmetric and   symmetric fission modes through  measurements of FFMDs from fission of $^{178}$Pt.

\section{Experiment}
\label{experiment}

$^{178}$Pt was  produced at the JAEA tandem accelerator \cite{jaea} in a complete fusion reaction $^{36}$Ar + $^{142}$Nd $\rightarrow$ $^{178}$Pt$^{*}$. A 75 $\mu$g/cm$^2$-thick $^{142}$Nd target was made by sputtering of
the $^{142}$NdF$_3$ material (isotopically enriched to 99$\%$) onto a thin (42 $\mu$g/cm$^2$) carbon backing facing the beam. Stability of the target performance with irradiation time was confirmed by the measurements of the $^{36}$Ar ions scattered  into a Si detector placed at backward angles, as well as by the constancy (within every beam energy setting) of the counting rate monitored during the experiment.
The $^{36}$Ar beam intensity was a few pnA, and the measurements were performed at three beam-energy settings (155, 170, and 180 MeV). 
Table~\ref{tab:table1} gives details on the energy balance of the formed CN. 
\begin{table*}[htp]
\centering
\caption{\label{tab:table1}Initial beam energy $E_{\rm beam}$ and its value (in brackets) at a mid-thickness of the target; mid-target CN excitation energy $E^{*}_{\rm CM}$ obtained from the reaction mass balance; average induced angular momentum $\bar{\ell}$; calculated fission-barrier height $B_{f,\bar{\ell}}$; average energy $\bar{E_{\nu}}$ taken away by pre-fission neutrons; rotational energy $E_{\rm rot}$; effective excitation energy $E^{\rm eff}_{\rm CM}$ derived as $E^{*}_{\rm CM}-B_{f,\bar{\ell}} - \bar{E_{\nu}} - E_{\rm rot}$;  TKE distribution components ${\rm TKE}^{\rm low}$ and ${\rm TKE}^{\rm high}$; and their widths $\sigma_{{\rm TKE}^{\rm low}}$ and $\sigma_{{\rm TKE}^{\rm high}}$. All values are in units of MeV, except for  $\bar{\ell}$ expressed in $\hbar$. The uncertainties are of statistical origin.}

\begin{threeparttable}

\begin{tabular}{clrrrrlcccc}
\hline 
$E_{\rm beam}$ & $E^{*}_{\rm CM}$ & $\bar{\ell}$\tnote{a}~~ 
& $B_{f,\bar{\ell}}$\tnote{b} 
&  $\bar{E_{\nu}}$\tnote{c} 
& $E_{\rm rot}$ & $E^{\rm eff}_{\rm CM}$ 
& ${\rm TKE}^{\rm low}$  & $\sigma_{{\rm TKE}^{\rm low}}$  
& ${\rm TKE}^{\rm high}$ & $\sigma_{{\rm TKE}^{\rm high}}$ \tabularnewline 
\hline \hline

155.0(153.9) & 38.6 &  9.0 & 12.7 & 0.3 & 0.7 & 24.9 
 & --- & --- & --- & --- \tabularnewline
170.0(168.8) & 50.5 & 28.2 & 10.1 & 9.9 & 5.0 & 25.5 
& 114.7(43) & 12.6(13) & 133.4(13) & 10.9(4) \tabularnewline
180.0(178.8) & 58.4 & 37.6 &  8.1 & 16.3 & 8.5 & 25.5 
& 114.6(64) & 15.4(16) & 131.2(9) & 12.6(3) \tabularnewline
\hline 

\end{tabular}

\begin{tablenotes}\footnotesize
\item[a] derived from the coupled-channel calculation of the CN production probabilities. \cite{Hagino1999}
\item[b] initial values from \cite{Moller2015Bf} corrected for rotation \cite{Sierk1986}.
\item[c] calculated in accordance with procedure described in \cite{Nishio2015}.
\end{tablenotes}
\end{threeparttable}
\end{table*}

The coincident fission fragments of $^{178}$Pt were detected with a two-arm time-of-flight (TOF) setup placed downstream the target, with two TOF arms positioned symmetrically at $\pm 60{^\circ}$ relative to the beam axis, with horizontal and vertical acceptance of $\pm 15{^\circ}$. 
The chosen detection angles allowed for similar angular acceptance for both mass-symmetric and mass-asymmetric fission events and, thus, excluded influence of the setup geometry on the observed fission properties.
Each TOF arm was comprised of a micro-channel plate based detector (MCP) and a position-sensitive multi-wire proportional counter (MWPC), providing the timing START and STOP signals. For the central trajectory, the  target--MCP foil distance was 67 mm, and the TOF base of 243 mm was identical for the two TOF arms. The MWPCs (active area of $200\times200$ mm$^2$) were operated with isobutane gas at a pressure of 1.5 Torr and had a 2 
$\mu$m aluminum-coated Mylar entrance window, whereas the MCP-based detectors were equipped with a thin (0.5\,$\mu$m) Mylar foil coated with Au and CsI (100{\AA} and 20{\AA} of thickness, respectively).
In addition to timing signals and spacial coordinates for the detected events, the MWPCs have also provided information on their partial energy loss in the isobutane.

\section{Results}
\label{results}

\begin{figure}[hbt]
\includegraphics[width=\linewidth]{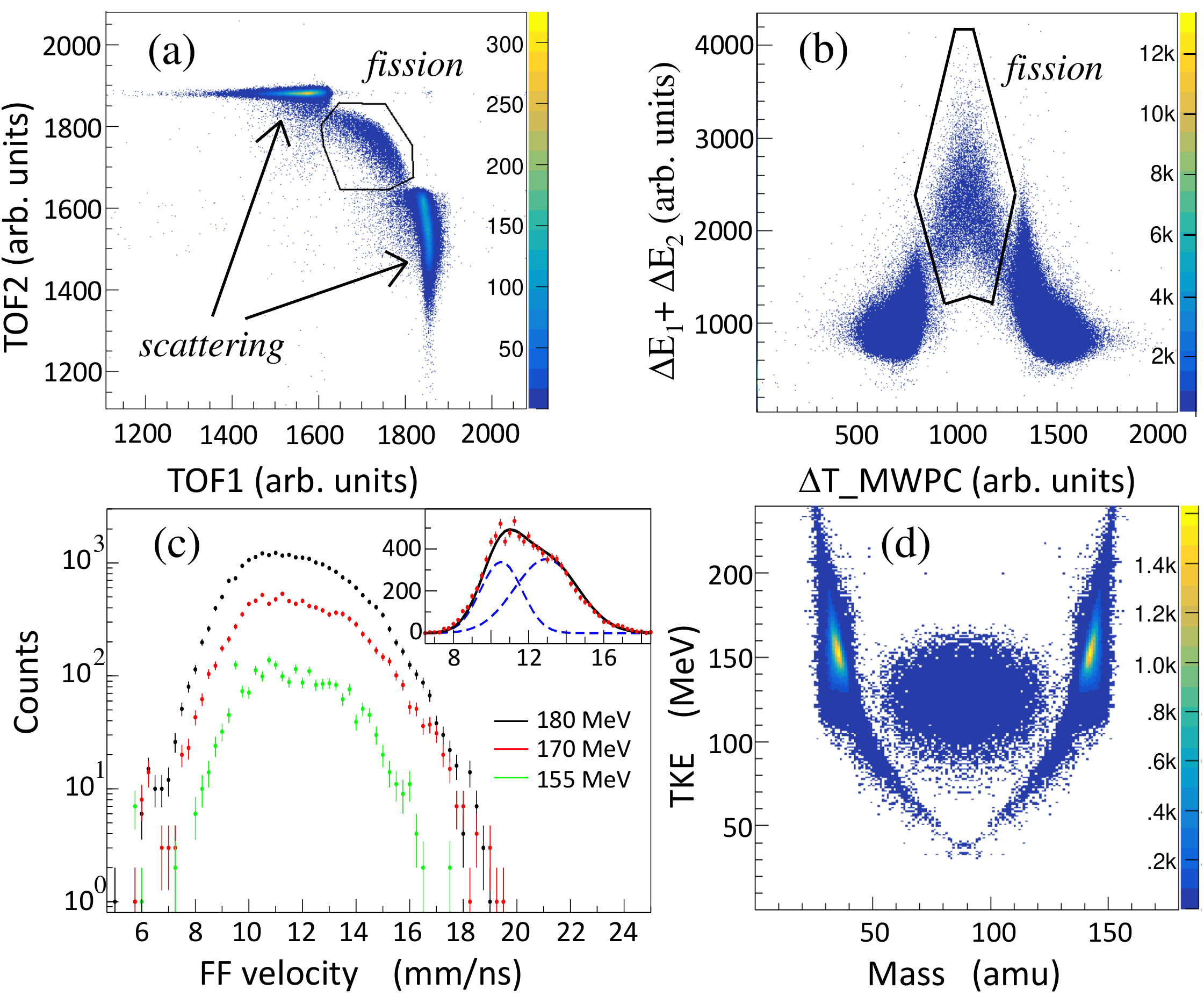}
\caption{
(a) Two-dimensional TOF1-TOF2 raw-data spectrum for coincident events at $E_{\rm beam}=170$ MeV. Events in the black contour are from fission; remaining events are from the projectile-target scattering. 
(b) Summed energy signals from the MWPC detectors for $E_{\rm beam}=170$ MeV plotted against the difference in their timing signals (coincidence with MCPs not demanded). Contoured are fission events producing more ionization in detectors, due to their significantly larger ionic charge states and, hence, higher effective charges that scattered beam/target nuclei.
(c) FF velocities after calibration with the scattered $^{36}$Ar beam, corrected for attenuation in the target and TOF detectors. The inset shows a typical (free) fit of the data at $E_{\rm beam}=170$ MeV.  
(d) Events from (a) represented in terms of their total kinetic energy and mass both calculated from experimentally obtained velocities, assuming fission process as the only events' origin. Group of events in the plot's center is coincident with data in contours in (a) and (b). 
}
\label{fig1}
\end{figure}

Figures~\ref{fig1}a-b give samples of recorded coincident data, in which experimental observables (timing signals and energies) are used without any preliminary treatment. Three groups of events are distinctly visible in the plots. Their identification as projectile/target scattering and fission events is obvious directly from the plotted raw data. 

For the follow-up analysis, we select fission events by making use of two conditions on the observables, indicated in Figs~\ref{fig1}a-b as contours
\footnote{
An alternative approach for the fission event selection is to extract from the measured data masses and total kinetic energies and to construct a corresponding correlation plot, as shown in Fig.~\ref{fig1}d. This analysis does not necessitate any prior gating but uses commont treatment  (two-body fission kinematics) for all of the measured events. The approach does not work if not fission but (elastic) scattering is assumed as the only origin of the detected events.  
}. 
 Angular information extracted from the MWPC impact coordinates (folding angles: see Ref.~\cite{Nishio2015} for details) was used to check for the selection quality.

For every identified fission event, velocities of coincident FFs were derived from the measured TOF values and TOF distances calculated with help of the MWPC coordinates. They were then calibrated with the scattered $^{36}$Ar beam and corrected for the reaction kinematics, as well as for attenuation in the target (calculated for a half of the tickness) and the TOF detectors' foils.

Figure~\ref{fig1}c shows the obtained FF velocities for one of the TOF arms.
The striking feature of these distributions is their pronounced non-symmetric character. A good description of the velocity spectra is achieved with a two-Gaussian fit, as demonstrated in the inset of Fig.~\ref{fig1}c. The asymmetry in the velocity spectra allows one to conclude that the fission of $^{178}$Pt produces fragments with different masses and is therefore predominantly asymmetric.

Importanly, the two-component velocity fits as in Fig.~\ref{fig1}c deliver very different distribution widths and thus do not yield the same integral for the expected light and  heavy fragment groups. This is a direct indication of presence of symmetric-fission events in the data.

\begin{figure*}[htb]
\includegraphics[width=\linewidth]{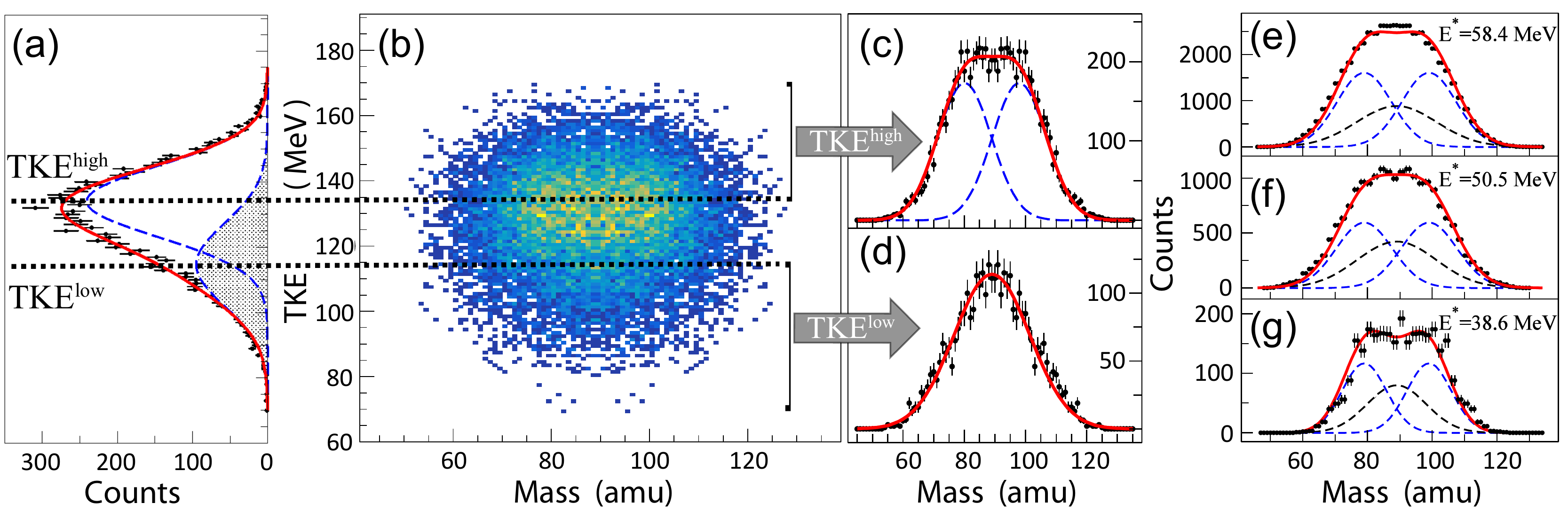}
\caption{(a) TKE distribution for $E_{\rm beam}=$170 MeV (projection of (b) onto the TKE-axis)  de-convoluted into two components
with derived positions of  ${\rm TKE}^{\rm high}$ and ${\rm TKE}^{\rm low}$, shown by dotted horizontal lines, see text for details. (b) TKE -- FF mass correlation obtained with events' selection as in Figs.~\ref{fig1}a-b.   TKE scale is identical for both (a) and (b).  Mass spectra gated on events above ${\rm TKE}^{\rm high}$ (c) and below ${\rm TKE}^{\rm low}$  (d)  fitted with a double- and single-Gaussian unconstrained function; fit results given by red lines. (e-g) Total FFMDs at different CN excitation energies (cf. Table~\ref{tab:table1}). 
Solid red lines result from a fit with fixed symmetric and asymmetric mode positions. Blue and black dashed lines show the asymmetric and symmetric fit components, respectively. Experimental mass resolution is $\sigma^{\rm exp}_{A}=2.9$\,amu, as deduced from the width of the $^{36}$Ar peak (not shown).}
\label{fig2}
\end{figure*}

The mass numbers $A_L$ and $A_H$ of light and heavy FF groups, respectively, along with their respective total kinetic energy (TKE), can be readily  derived from the fragments' velocities $v_L$ and $v_H$, under assumption of no particle emission (i.e., $A_L+A_H=A_{\rm CM}$) from the compound nucleus $A_{\rm CM}$ during the pre-fission stage:  $A_L v_L=A_H v_H$ and ${\rm TKE}=0.5 A_{\rm CM} v_L v_H$. 
An example of the deduced TKE-mass data is shown in Fig.~\ref{fig2}b. 
Projection of the data in Fig.~\ref{fig2}b on the TKE-axis gives the TKE distribution (Fig.~\ref{fig2}a), 
whose average value $\overline{\rm TKE}$ and width $\sigma_{\rm TKE}$ are found to slightly change with the increasing beam energy ($\Delta\overline{\rm TKE}=-1.9(2)$\,MeV, $\Delta\sigma_{\rm TKE}=1.2(2)$ \,MeV for the measured $E_{\rm beam}$ range). This corraborates recent results on the TKE parameters' behaviour in $^{180,190}$Hg \citep{Nishio2015} and is generaly inline with positive and negative slopes in $\frac{d\overline{\rm TKE}}{dE^*_{CN}}$ and $\frac{d\sigma_{\rm TKE}}{dE^*_{CN}}$, respectively, known for actinides (cf., e.g., \cite{Straede1987}). 

The TKE distribution in Fig.~\ref{fig2}a is clearly skewed. 
The simulated FF energy straggling in the target and TOF detectors' foils could not reproduce the observed asymmetry effect in the TKE, unless unrealistic assumptions  are made about the inhomogeneity of the MCP foil (thickness varying from zero till 10 times the nominal value of 0.5$\mu$m).
 Similarly-skewed TKE distributions were obtained also at $E_{\rm beam}$=155 and 180\,MeV. 
Based on the velocity analysis, an unconstrained two-Gaussian fit was carried out to describe the TKE data. 
This fit, statistically reliable only at the two higher energies, yields two TKE components placed at ${\rm TKE}^{\rm low}$ (maximum of the shadowed-area curve in Fig.~\ref{fig2}a) and ${\rm TKE}^{\rm high}$ (maximum of the other dashed curve); their numerical values are given in Table~\ref{tab:table1}.

The TKE components ${\rm TKE}^{\rm low}$ and ${\rm TKE}^{\rm high}$ are linked to the symmetric and asymmetric fission modes. This is demonstrated by the difference in the shape of the partial MDs 
constructed with events in Fig.~\ref{fig2}b in the regions below ${\rm TKE}^{\rm low}$ and above ${\rm TKE}^{\rm high}$ and projected on the mass-axis (cf. the dotted lines and arrows in the Figure): narow and clearly symmetric in Fig.~\ref{fig2}d and wide and flat-top in Fig.~\ref{fig2}c. 
The best-fit desciptions of partial MDs in Figs~~\ref{fig2}c-d are achieved with one- and two Gaussians, respectively. The latter determines the light ($A_{L}$=79(1) amu) and heavy ($A_{H}$=99(1) amu) FF peak positions. 
Thus, our experimental results shown in Fig.~\ref{fig2}c-d offer the first direct experimental evidence of the co-existing symmetric and asymmetric fission modes in the $^{178}$Pt nucleus and in the sub-lead region. Contrary to the Mulgin {\it et al.} \cite{Mulgin1998} who interpreted earlier experimental data close to the $\beta$-stability line around A$\sim$200 \cite{Itkis1985,Itkis1991} within a liquid-drop model with phenomenological shell corrections added, our conclusion on the coexistence of two modes in $^{178}$Pt is based on the assumption-free deconvolution of  experimental TKE-mass data which makes the result unambiguous. 

The experimental total FFMDs are shown in Figs.~\ref{fig2}e-g by the black circles.
Solid red and dashed lines are results of the analysis in terms of two fission modes, with the fit function composed of three Gaussians with fixed positions as obtained above.  
Overall, a good description of the experimental data is achieved.
The asymmetric mode is found to be dominant, in accordance with the velocity analysis. The weight of the symmetric mode  amounts to $\sim$31$\%$ at the three considered beam energies.
Thus, in contrast to actinides \cite{Straede1987}, the balance between symmetric and asymmetric modes in the FFMDs does not seem to be significantly affected by the excitation energy.     
This can be explained in terms of the energy considerations of Table\,\ref{tab:table1}: corrections to the excitation energy $E^{*}_{\rm CM}$ due to possible neutron emission\footnote{
Proton emission has been neglected as it affects less than 10\% of fission events at the highest excitation energy, as estimated with the statistical code GEF \cite{SCHMIDT2016107}
} 
$\bar{E_{\nu}}$, rotational energy $E_{\rm rot}$ of the CN and the rotation-dependent fission-barrier height $B_{f,\bar{\ell}}$ reduce the initial spread of 20\,MeV in $E^{*}_{\rm CM}$, 
resulting in practically identical ($\sim$25 MeV) effective excitation energy $E^{\rm eff}_{\rm CM}$.

\section{Interpretation}
\label{interpretation}

\begin{figure}[h]
\includegraphics[width=\linewidth]{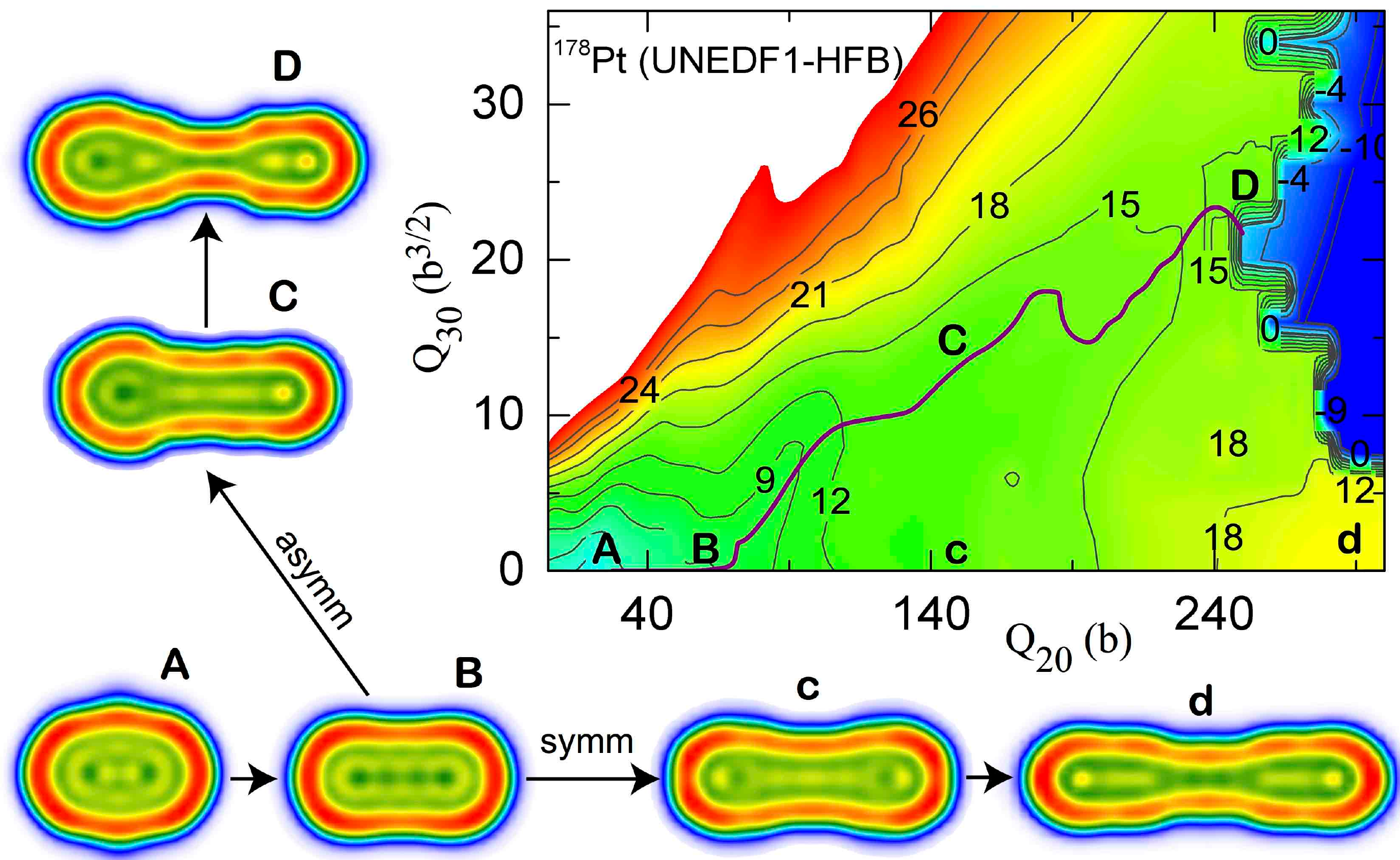}
\caption{PES of $^{178}$Pt in the ($Q_{20}$, $Q_{30}$) plane  calculated in UNEDF-HFB. The solid thick line  indicates the static fission path obtained by the local minimization of PES. To illustrate the shapes on the way to fission, and  the emergent pre-fragments, 
the neutron localization functions~\cite{Zhang2016,Sadhukhan2017} corresponding to various intrinsic configurations along the asymmetric (ABCD) and symmetric (ABcd) paths  are plotted.}
\label{UNEDF} 
\end{figure}

\begin{figure}[h]
\includegraphics[width=\linewidth]{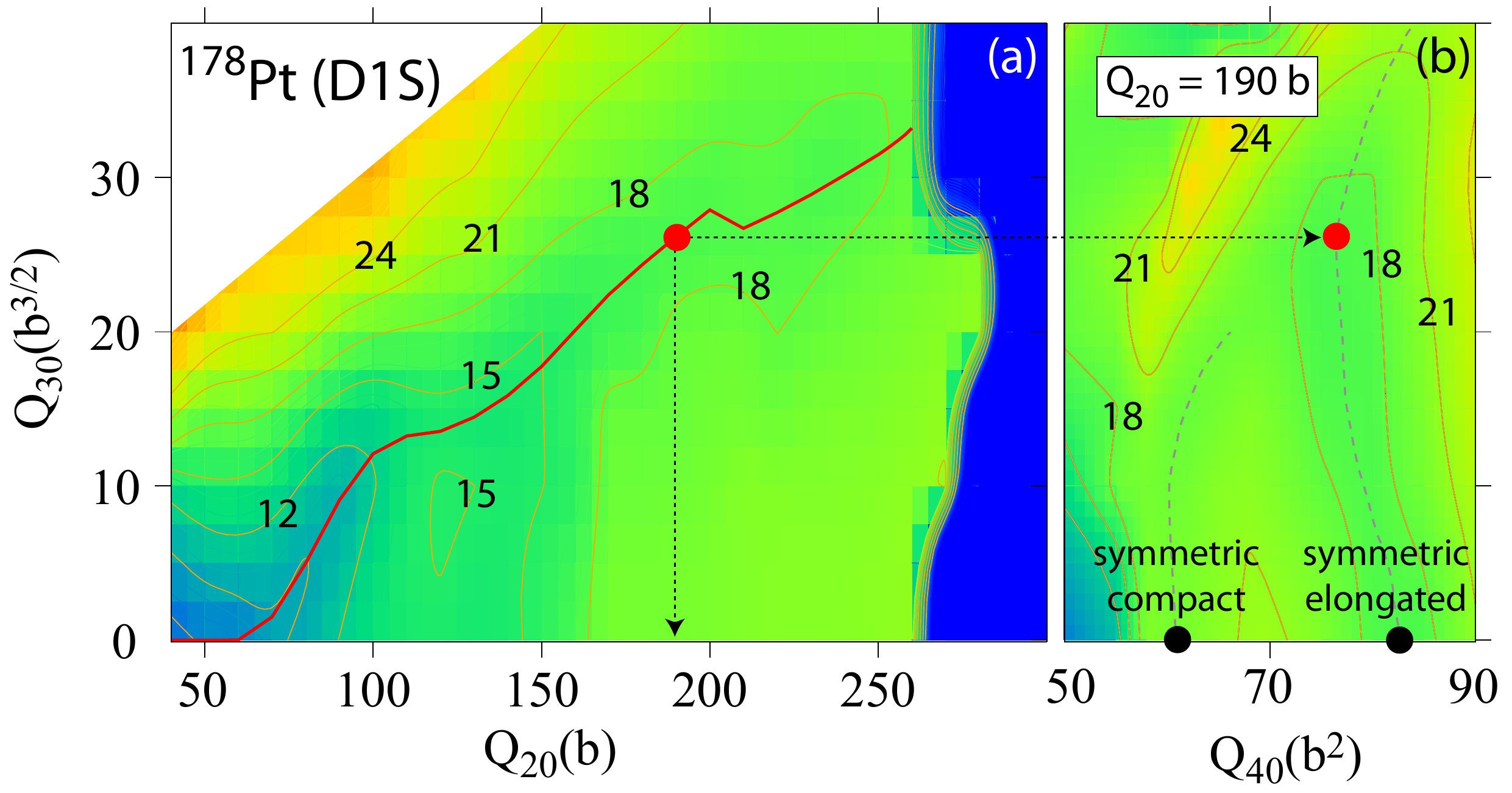}
\caption{PES of $^{178}$Pt in the ($Q_{20}$, $Q_{30}$) plane (a)
and in the ($Q_{30}$, $Q_{40}$) plane at $Q_{20}=190$\,b (b) obtained in D1S. The solid thick line in (a) indicates the static fission path obtained by the local minimization of PES. Dashed lines in (b) indicate the symmetric PESs corresponding to compact (smaller $Q_{40}$) and elongated (larger $Q_{40}$) fragments. The minimum corresponding to the static fission path in (a) is marked by the red dot.}
\label{D1S}
\end{figure}

To interpret experimental results,  nuclear density functional theory (DFT) calculations have been performed within two Hartree-Fock-Bogolyubov  frameworks employing the
Skyrme  UNEDF1-HFB \cite{schunck2015error} and Gogny D1S \cite{Berger1991} energy density functionals (cf. Figs~\ref{UNEDF} and \ref{D1S}, respectively). The constrained calculations were performed in the collective space of quadrupole 
($Q_{20}$) and octupole ($Q_{30}$) moments, and also in the hexadecapole direction $Q_{40}$ in the D1S model.
It is encouraging to see that
both approaches yield very similar picture of PES. In both calculations, the static fission path leads to the mass-asymmetric $A_{L}/A_{H}\approx$80/98 split, which matches the experimental result very well.
(We note
that the Brownian shape-motion method Ref.~\cite{Moller2015} predicts a strongly asymmetric split with $A_{L}/A_{H}\approx$70/108  at the CN excitation energy of 16.5\,MeV.)

To understand the formation of fragments corresponding to the $^{178}$Pt fission pathways, we use the concept of nucleon localization functions (NLFs)~\cite{Zhang2016}. Within this framework, the elongated configurations on the way to scission are composed of two clusters (pre-fragments) connected by a neck. At scission, the neck nucleons are redistributed into pre-fragments, producing the final fission fragments. 
As shown in Ref.~\cite{Sadhukhan2017},
NLFs quantify the appearance of pre-fragments more efficiently than
nucleonic density distributions as the concentric patterns in NLFs -- due to shell structure in the nuclear interior -- are averaged out in density distributions.
Figure~\ref{UNEDF} displays the resulting NLFs along the two fission pathways: asymmetric (ABCD) and symmetric (ABcd). 
Based on the analysis of NLFs according to the procedure of Ref.~\cite{Sadhukhan2017}, the asymmetric pre-scission  configurations marked ``C"  and ``D" in Fig.~\ref{UNEDF}  are composed of a nearly-spherical cluster around  $^{86}$Sr and a lighter deformed pre-fragment. Such a structure  results  in FFs around $^{98}$Mo and $^{80}$Kr. As far as the symmetric configuration ``c" is concerned, its pre-fragments can be associated with spherical $^{64}$Ni nuclei.

The static fission valley in Figs.~\ref{UNEDF} and \ref{D1S} evolves on a fairly flat landscape, in contrast to a typical situation in heavy actinides (see e.g. \cite{McDonnell2014,Ichikawa2012}). Absence of any ridge in the area of  low octupole moments, along with a fairly small energy difference between the asymmetric and  symmetric paths, suggests a possibility for a competition between different fission modes.
At present, a  detailed description of this competition is difficult to assess theoretically, as the  post-scission configurations associated with fusion valleys \cite{Warda2012}
enter the picture and produce a sudden drop in PES at very large elongations (cf. Figs.~\ref{UNEDF} and \ref{D1S}a), which makes it practically impossible to follow adiabatically the original fission trajectory. 

A detailed analysis of the PES in Fig.~\ref{D1S}b shows that the plateau predicted for nearly-symmetric shapes around $Q_{20}= 190$\,b in the region between the paths CD and cd, has  a rather complicated structure. Namely, at the same values of  quadrupole and octupole moments, two local symmetric PES minima with similar energies but distinct hexadecapole moments and nuclear density distributions are  found. One of these solutions, with  $Q_{40}\sim 60$\,b$^2$, corresponds to  compact fragments, while that with $Q_{40}\approx 85$\,b$^2$ can be associated with very elongated fragments. In both models, the symmetric pathway associated with elongated-fragment configurations, expected to have lower TKE,  is predicted to be energetically slightly more favored than that associated with compact fragments. Therefore, it cannot be excluded that the symmetric fission mode seen experimentally contains contributions from both structures. 
It is interesting to see that competing fission pathways involving similarly asymmetric, compact, or elongated shapes have been predicted for multimodally fissioning nuclei in the fermium region \cite{Warda2002,Staszczak2009}, i.e., for nuclei with much larger values of $A_{CN}$ and $N/Z$.

Experimentally, we find that both symmetric and asymmetric fission modes follow the trend previously observed in  heavier, trans-lead, nuclei \cite{Zhao1999}. In particular, higher values of TKE in the asymmetric mode (cf. Table~\ref{tab:table1}) -- which also match well the TKE=135.9\,MeV value expected from the Viola systematics \cite{Viola1985} -- are indicative of less deformed scission configurations, whereas for the symmetric mode, highly elongated FF shapes are expected from its lower TKE values. 
This finding is consistent with the  shapes  of nucleon localizations shown in Fig.~\ref{UNEDF}:  symmetric configuration ``d" 
corresponds to highly deformed  fragments  
without a well defined  neck. As discussed above, a  similar configuration associated with
symmetric elongated fragments has been predicted  in the D1S model: in Fig.\,\ref{D1S}b it is marked by  a black dot at  $Q_{40}\approx 85$\,b$^2$ and $Q_{30}\approx 0$.

\section{Conclusions}
\label{conclusions}

In summary, the FFMDs of $^{178}$Pt produced in a complete fusion reaction $^{36}$Ar + $^{142}$Nd are found to be predominantly asymmetric, with the most probable mass division $A_{L}\thickapprox{79}$ and $A_{H}\thickapprox{99}$. The combined analysis of the FFMDs and TKE distributions made it possible to separate  asymmetric and symmetric fission modes. 
It is found that the asymmetric mode is associated with larger TKE values than the symmetric mode. Moreover, its average TKE follows the systematics \cite{Viola1985} established for nuclei with $N/Z\sim 1.5$, which suggests the asymmetric mode's insensitivity to the isospin of the CN, at least for $A_{CN}>177$.  

The UNEDF1-HFB and D1S calculations support the exprimental results. Namely, they correctly reproduce the measured mass division associated with the dominant asymmetric fission mode, and they predict highly elongated pre-scission configurations along the symmetric fission path, which is in accordance with the lower experimental TKE value for this mode. 

The present work provides new experimental information on the extension of the recently-discovered island of asymmetric fission towards lower atomic numbers.  
For the first time,  the interplay between different fission modes has been  found in a nucleus from the sub-lead region.  The result  provides strong motivation for extending microscopic models of fission to 
 FFMDs and TKE distributions at nonzero excitation energies. Finally,  beyond-DFT extensions of the current formalism are needed, as the PESs predicted for pre-lead nuclei are generally very flat in the pre-scission region, resulting in possible interferences between asymmetric and symmetric fission modes.

\section*{Acknowledgements}

The authors express their gratitude to the JAEA tandem crew for the help in performing the $^{178}$Pt experiment and to the GSI target group for making the $^{142}$Nd target. 
This work was in part supported by the JAEA Reimei and STFC (UK) grants;  by the
U.S. Department of Energy under Awards No. DE-NA0002847
(NNSA, the Stewardship Science Academic Alliances
Program) and No. DE-SC0018083 (Office of Science,
Office of Nuclear Physics NUCLEI SciDAC-4 Collaboration); and by the
Polish National Science Centre under Contract
No. 2016/21/B/ST2/01227.





\section*{References}

\bibliographystyle{elsarticle-num}
\bibliography{references}







\end{document}